\documentclass{article}  
\usepackage{breckenr2005}
\usepackage{subfigure}
\usepackage{floatflt}
\usepackage{graphicx}
\usepackage{setspace}
\frompage{000} \topage{000}                                              
\def\rightmark{High $p_T$ spectra from STAR}
\def\leftmark{O. Barannikova}
 
\title{High transverse momentum identified particle spectra in 200 GeV collisions from STAR} 
\authors{
{Olga Barannikova$^1$ \\%
\it{(for the STAR collaboration)} }\\[2.812mm]
{\normalsize
\hspace*{-8pt}$^1$ Purdue University, \\ 
West Lafayette, IN, USA\\[0.2ex] 
}}
 
\abstract{Significant baryon over meson enhancement was measured at RHIC in the intermediate transverse momentum range of $p_T=2-4$ GeV/$c$ ("baryon-meson puzzle"). With STAR detector we were able to extend particle identification towards higher transverse momentum offering further insights into the particle production mechanisms at intermediate to high $p_T$. In this paper we present results on charged pion, proton and anti-proton  spectra and ratios at intermediate to high $p_T$ exploiting the relativistic rise of the specific ionization energy loss measured in the STAR Time Projection Chamber.
These measurements provide valuable information about the production mechanisms of particles at intermediate $p_T$ in relativistic heavy ion collisions, e.g. coalescence/recombination versus jet fragmentation.  }


\begin{document}
 
\maketitle
\setcounter{page}{1}

\section{Introduction}\label{intro}

Ultra-relativistic heavy ion collisions provide unique environment to study particle production mechanisms in the matter under extreme conditions of high temperature and energy densities. Different mechanisms govern hadron formation in the different kinematic regions: it is believed that at large transverse momenta, particle production in ultra-relativistic heavy ion collisions is dominated by jet fragmentation, and at low momenta by soft particle production, well described by hydrodynamical models \cite{matt5}. Studies of particle production in the intermediate transverse momentum range ($2<p_T<4$~GeV/$c$) from Relativistic Heavy Ion Collider yielded puzzling results, that show significant baryon over meson enhancement ("baryon-meson puzzle"). Reported proton to pion ratio of unity\cite{phenix} is significantly larger than that observed  in elementary collisions\cite{matt7}.  Baryon-antibaryon ratio was also reported constant up to at least $p_T=4$~GeV/$c$ \cite{phenix,matt}, contradicting pQCD calculations that predict a decrease in such ratios \cite{matt2}.
Understanding the mechanisms of particle production in that intermediate momentum range and at higher momentum is a main physics motivation for the analysis presented in this paper. We present first STAR results on extended identification for charged pions, protons and antiprotons for the transverse momentum range of $2.5 < p_T < 7$GeV/$c$.

\section{Data analysis}\label{techno}

The measurement reported here was carried out with the data sample of central trigger Au+Au events collected at the center-of-mass energy of 200 GeV  in the Run-2 at the Relativistic Heavy Ion Collider (RHIC) by the STAR (So\-le\-noidal Tracker at RHIC) experiment. The STAR detector \cite{TPC} consists of several detector subsystems in a large so\-le\-noid magnet, including a time projection chamber (TPC), a scintillator barrel (CTB), and two zero degree ca\-lo\-rimeters (ZDC) \cite{pbar9}. The magnet was operated at 0.5 T. CTB  measuring the produced charged particle multiplicity around mid-rapidity and ZDC essentially measuring neutral spectator energy were used for triggering.

In this analysis the coincidence of the two ZDCs  with an addition of a high CTB signal, provided the collision trigger. 
In addition to the hardware trigger, in the off-line analysis the collision centrality was
determined from the measured charged particle multiplicity at mid-rapidity from the minimum bias sample. Thus the data sample corresponding to 5\% most central Au+Au events was selected.  In this analysis the total of 0.5~M events for this centrality bin was selected after requiring the primary interaction point to be within 25~cm of the TPC center to insure uniform acceptance and good detector resolution.

\subsection{Analysis technique}\label{details}

The analysis was carried out with the TPC measurements. TPC advantage is its  large, uniform acceptance and capability of measuring and identifying a substantial fraction of the particles produced in heavy ion collisions. In our previous publications specific ionization energy loss in the TPC material was successfully used for identification of charged pions and kaons up to  $p_T \simeq 0.7$ GeV/$c$, and protons and antiprotons up to  $p_T \simeq 1.2$ GeV/$c$ \cite{pbar,p,k,pi,PRL}. Those momentum values are usually quoted as limiting for particle identification with STAR TPC. However, STAR identification capability with dE/dx technique can be further extended into the relativistic rise region of the energy loss momentum dependence.

\begin{figure*}[hbt]
\centerline{\epsfxsize=0.6\textwidth\epsfbox[0 0 567 449]{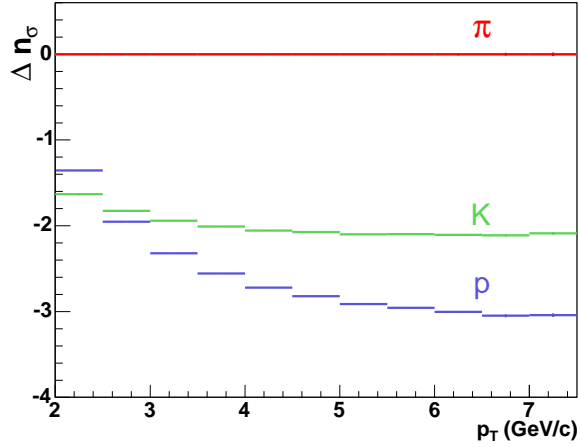}}
\caption{ Relative separation  between kaon (green) and proton (blue) $dE/dx$ bands with respect to the expected position of the pion distribution (red) in units of  $\sigma_{\pi}$.}
\end{figure*}

At the transverse momentum above $2.5$GeV/$c$,  there is a substantial difference in the ionization energy loss distribution between pions and other hadrons due to the pion relativistic rise of the $dE/dx$.  With the TPC  $dE/dx$ resolution of $\sim7-8\%$, this allows to identify pions from other hadrons above this momentum  by the TPC at the level of 2 standard deviations.   
Figure 1 shows the relative distances  between kaon and proton $dE/dx$ bands with respect to the expected center position of the pion distribution (plotted at zero).  
This relative distance is defined in terms of $n_{\sigma_{\pi}}$,  the normalized $dE/dx$ of pions:

$$n_{\sigma_{\pi}}^{X}=log((dE/dx) / B_{\pi})/\sigma_{\pi} -  log((dE/dx)/B_{X} )/ \sigma_{\pi},$$
where $X$ can be $e^{\pm},K^{\pm}$ or $p(\bar{p})$. $B_{X}$ is the theoretical expectation for $dE/dx$ for particle $X$, and $\sigma_{\pi}$ is the $dE/dx$ resolution of TPC. The observed behavior is in agreement with theoretical expectations from Bichsel function calculations of the energy loss in the TPC material\cite{TPC}.  

\begin{figure*}[hbt]
\centerline{\epsfxsize=0.45\textwidth\epsfbox[0 0 567 449]{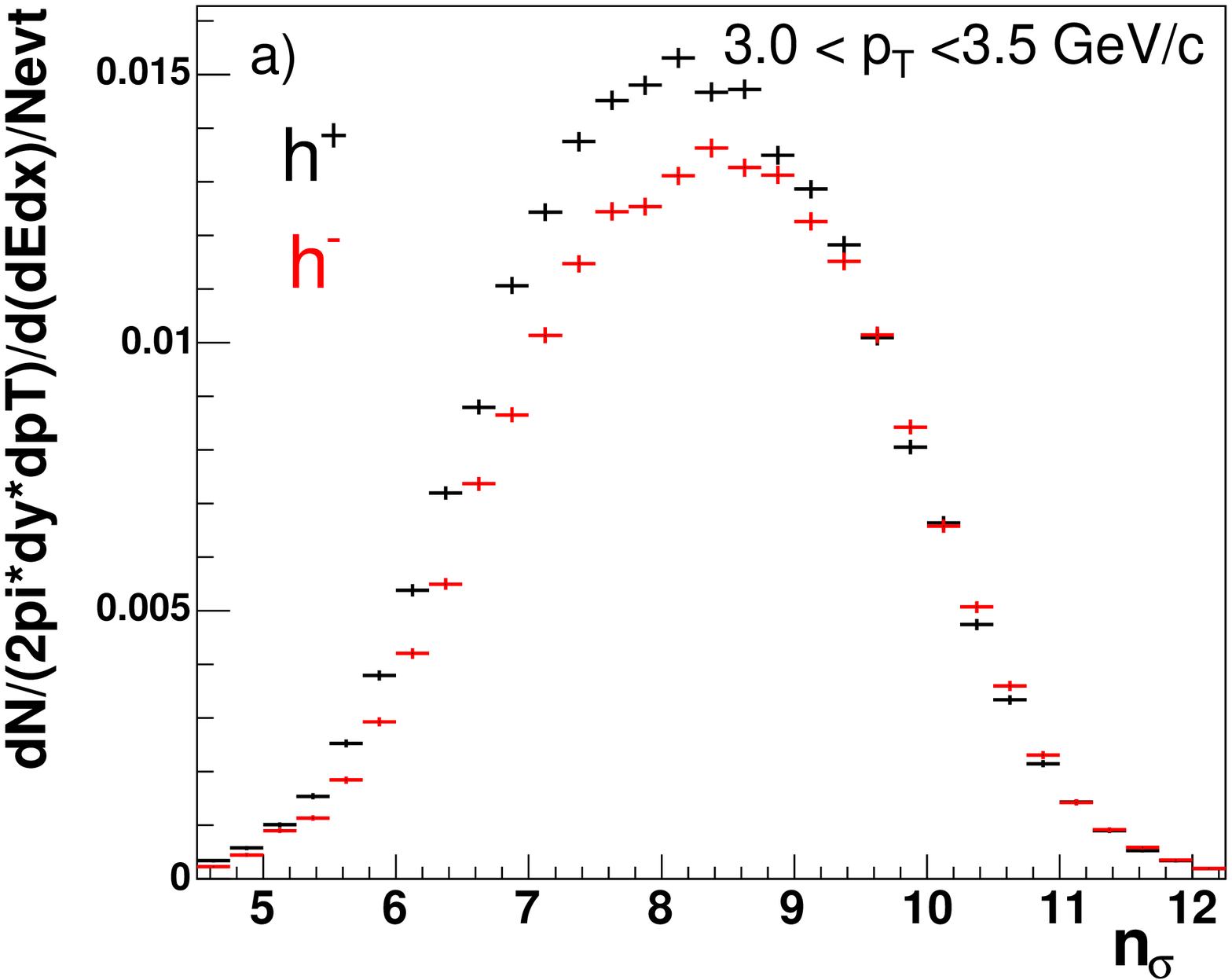}
\epsfxsize=0.45\textwidth\epsfbox[0 0 567 449]{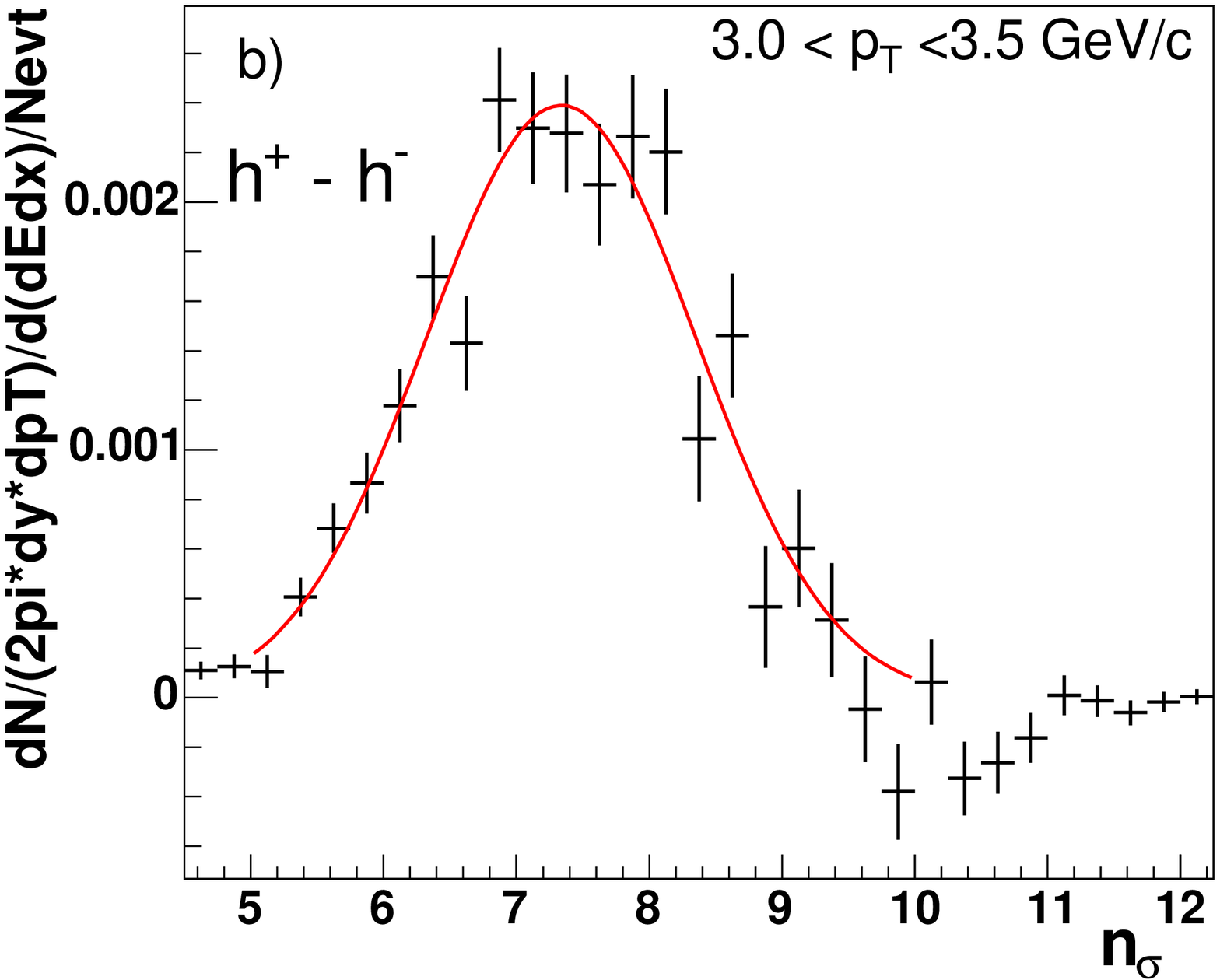}}
\caption{ Distribution of $n_{\sigma_{\pi}} + 10$ for positive (black) and negative (red) charged hadrons (a) and their difference $h^{+}-h^{-}$ (b) for the momentum range of (3.0,3.5)GeV/$c$.}
\end{figure*}

The $n_{\sigma_{\pi}}^{\pi}$ distribution is a normal Gaussian distribution with an ideal calibration. For plotting reason we displaced the $n_{\sigma_{\pi}}$ by 10 in Figure 2$a$.
The difference between negative and positive hadrons $n_{\sigma_{\pi}}$ distributions, evident from Figure 2, is expected to be dominated by net-protons present at mid-rapidity: the yield difference between positive and negative inclusive charged hadrons is $h^{+}-h^{-}=(p-\bar{p})+(K^{+}-K^{-})+(\pi^{+}-\pi^{-}) \simeq (p-\bar{p})$). This expectation is also confirmed by our measurements from STAR Time-of-Flight (TOF) detector \cite{TOFHQ2004}. We use the peak positions of $dE/dx$ distribution from $h^{+}-h^{-}$ (Figure 2$b$) as our initial estimate of that of proton, which provide additional data-driven  handle on proton-pion separation.  

Figure 1 also shows that the $dE/dx$ separation between kaon and proton is one $\sigma$ or less for the transverse momentum range studied, which is insufficient for proton-kaon identification. We use neutral kaon measurements performed by STAR\cite{matt13} and $K^+/K^-$ ratio from our low momentum $dE/dx$ measurement from the same data sample  to estimate charged kaon yield and additionally constrain the fit to solve proton-kaon unambiguity.

 Figure 3  shows ratio of the negative hadrons over positive hadrons as a function of $dE/dx$ in units of $n_{\sigma_{\pi}}$ at $3.0<p_T<3.5$GeV/$c$. Since the energy loss of particles in TPC is independent of its charge, the
dependence of $h^{-}/h^{+}$ on $n_{\sigma_{\pi}}$ is due to changing  particle composition. Two distinct plateau regions seen in  Fig.3 provide a simple way to estimate raw $\bar{p}/p$ and $\pi^{-}/\pi^{+}$ ratios. Particle-antiparticle ratios obtained from those plateau regions are used as systematic cross-check for the ratios extracted from final multi-gaussian fit described below.

\begin{figure*}[hbt]
\centerline{\epsfxsize=0.6\textwidth\epsfbox[0 0 567 449]{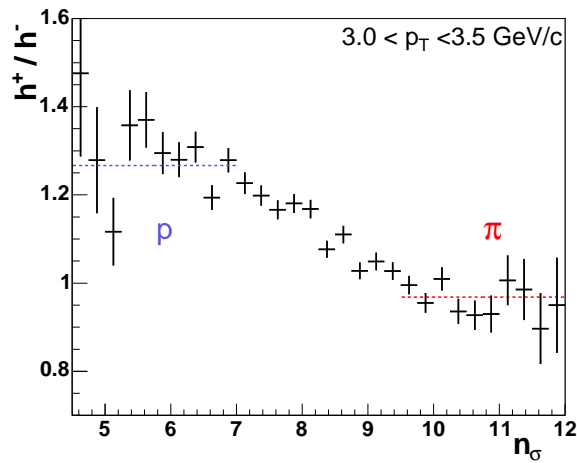}}
\caption{$h^{-}/h^{+}$ as a function of $dE/dx$ in units of $n_{\sigma_{\pi}}$ at $3.0<p_T<3.5$GeV/$c$.}
\end{figure*}

In order to statistically identify char\-ged pions, protons and antiprotons and extract their yields, for each momentum bin  we fit $n_{\sigma_{\pi}}$ distributions of positive and negative hadrons simultaneously with the sum of six Gaussian distributions. Figure 4$a$ shows fit results for one of the momentum bins.  The width is a free parameter and the same for  all six Gaussians. Centroid positions for positive and negative sings of the same specie are the same.
The relative distances of  $n_{\sigma_{\pi}}^{K}$ and $n_{\sigma_{\pi}}^{p(\bar{p})}$ are fixed to be proportional to that of the theoretical distribution and are scaled with the centroid values  of the pion and proton Gaussians, which are free parameters.

Charged pion, proton and antiproton spectra has been obtained from the raw particle yields extracted from the multi-gaussian fits and corrected for detector acceptance and tracking efficiency. Figure 4$b$ shows spectra for negative particles  (pions shown in filled red symbols, antiprotons - in filled blue symbols). In the same picture open symbols (respective colors) shows STAR low momentum  measurement for charged pions, obtained with TPC $dE/dx$ technique \cite{PRL} and antiprotons measured with TPC (via $dE/dx$) and RIHC detectors \cite{matt}. Black symbols shows STAR charged hadron measurement \cite{highpt}.  

\begin{figure*}[hbt]
\centerline{\epsfxsize=0.45\textwidth\epsfbox[0 0 567 449]{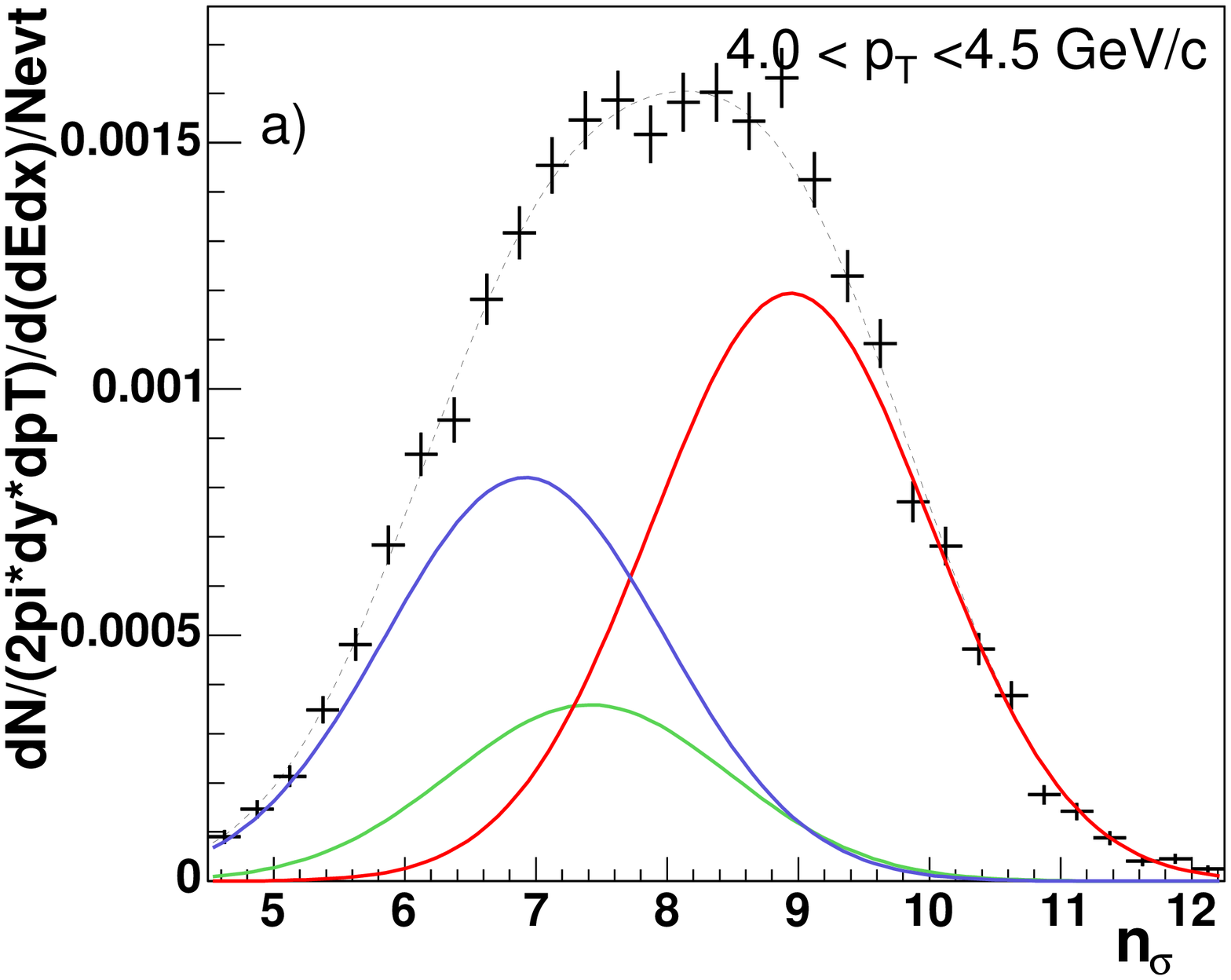}
\epsfxsize=0.45\textwidth\epsfbox[0 0 567 449]{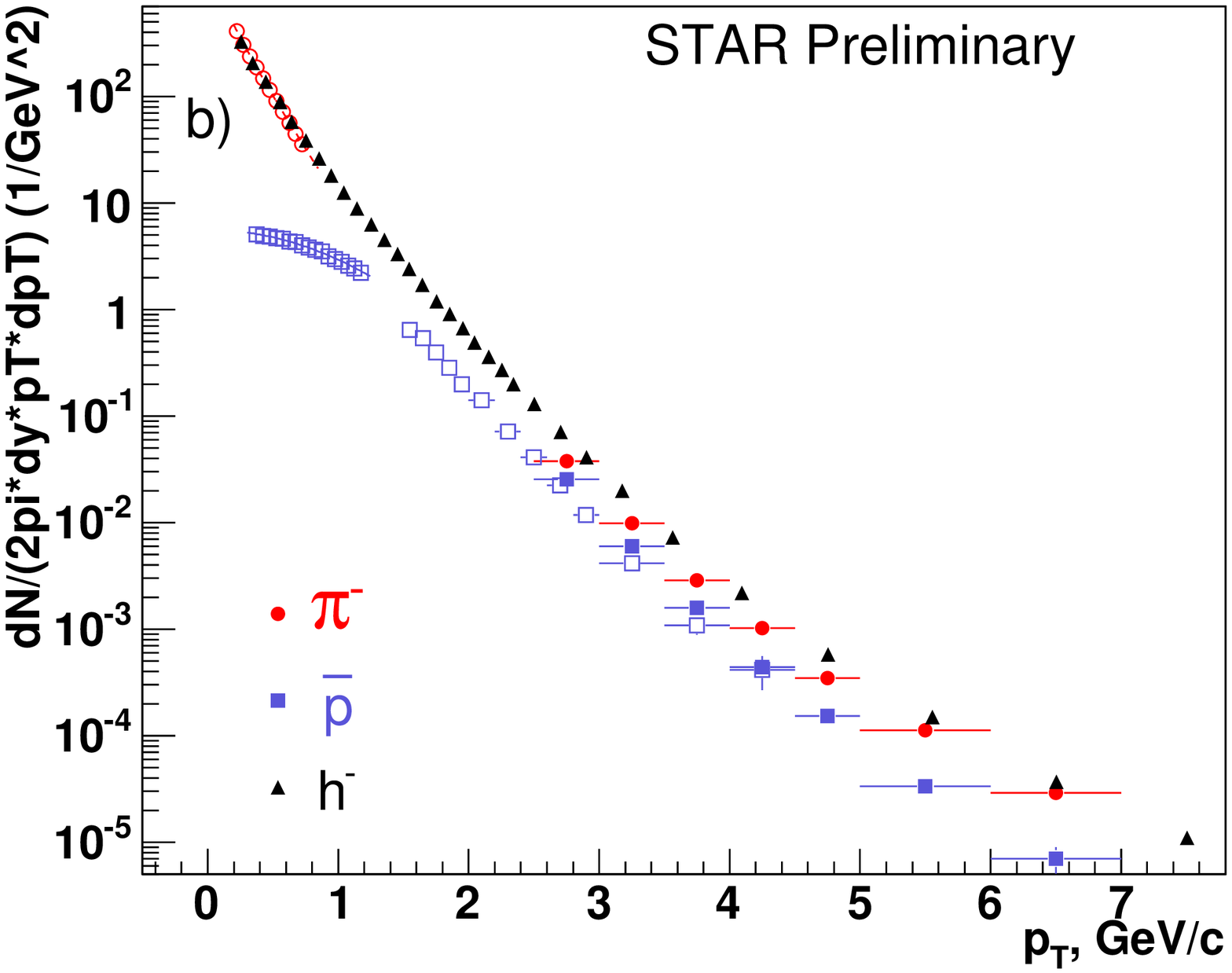}}
\caption{ a) $n_{\sigma_{\pi}}+10$ for positive hadrons in the momentum range of (4.0,4.5)GeV/$c$ fitted with multi-gaussian function defined in the text. Black line shows the actual fit result, results for pion (red), proton (blue) and kaon (green) contributions are shown in colors. b) Identified spectra for negative  pions (filled red symbols) and  antiprotons (filled blue symbols).Open symbols (respective colors) show STAR low momentum  measurement for the same particles. Total charged hadron distribution is shown in black symbol for reference. }
\end{figure*}

\section{Results}

 With the measurements of charged pion, protons and antiprotons spectra, presented above, we have studied transverse  momentum dependence of relative particle production in central Au+Au collisions at RHIC. Particle-antiparticle ratios as function of $p_T$ are shown in Figure 5. Corresponding measurements performed with RICH detector at lower momentum are also shown for comparison. $\pi^-/\pi^+$ ratio (Fig. 5$a$) is consistent with unity and independent of transverse momentum. $\bar{p}/p$ measurements at high transverse momentum are similar to those at low $p_T$ \cite{pbarp} with possible indication of slight decrease in the ratio above 3 GeV. Even if the decrease is taken seriously, our observation contradicts  to pQCD inspired models which predict a stronger decrease in the ratio over all $p_T$ \cite{b26,b27}.

\begin{figure*}[hbt]
\centerline{\epsfxsize=0.45\textwidth\epsfbox[0 0 567 449]{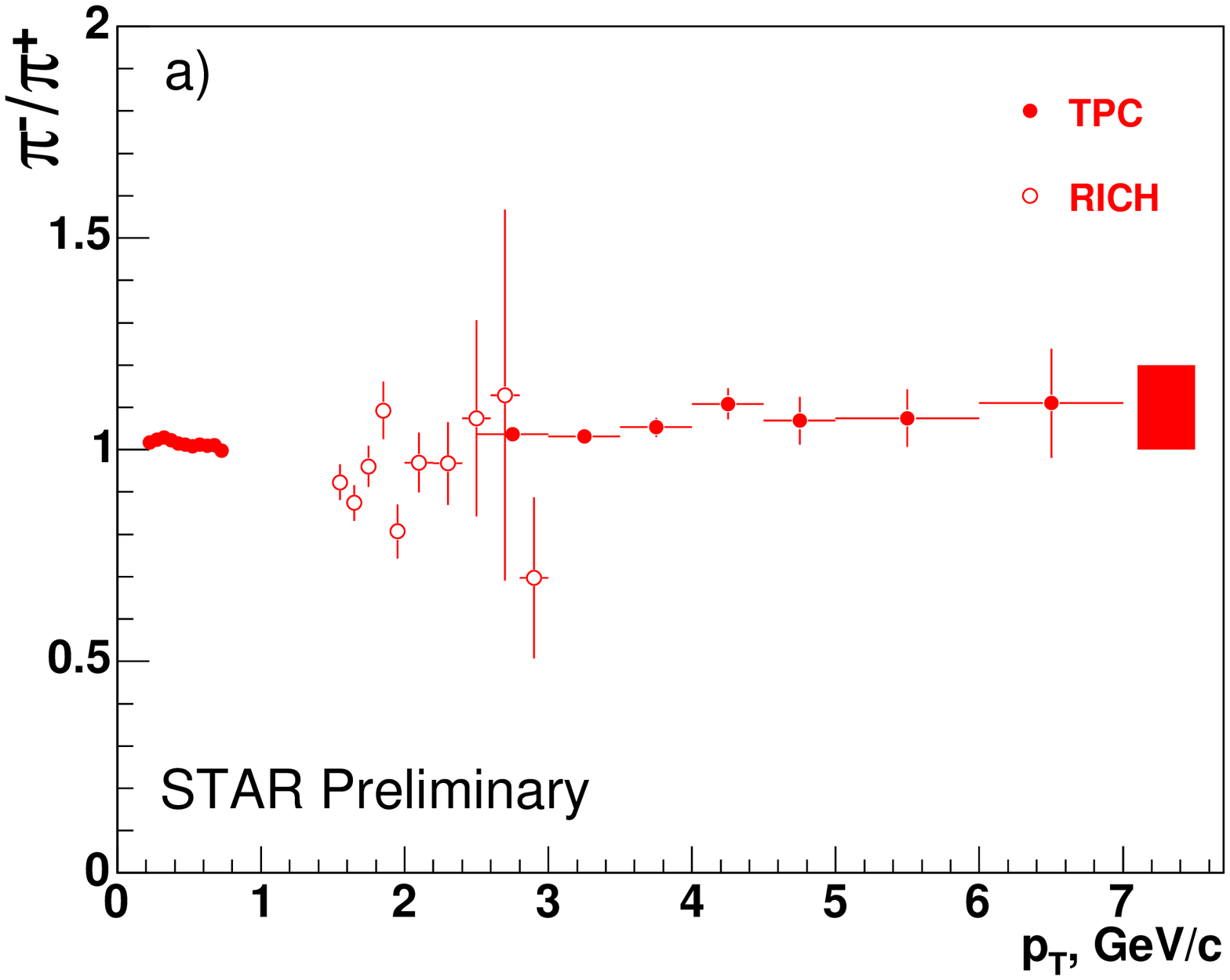}
\epsfxsize=0.45\textwidth\epsfbox[0 0 567 449]{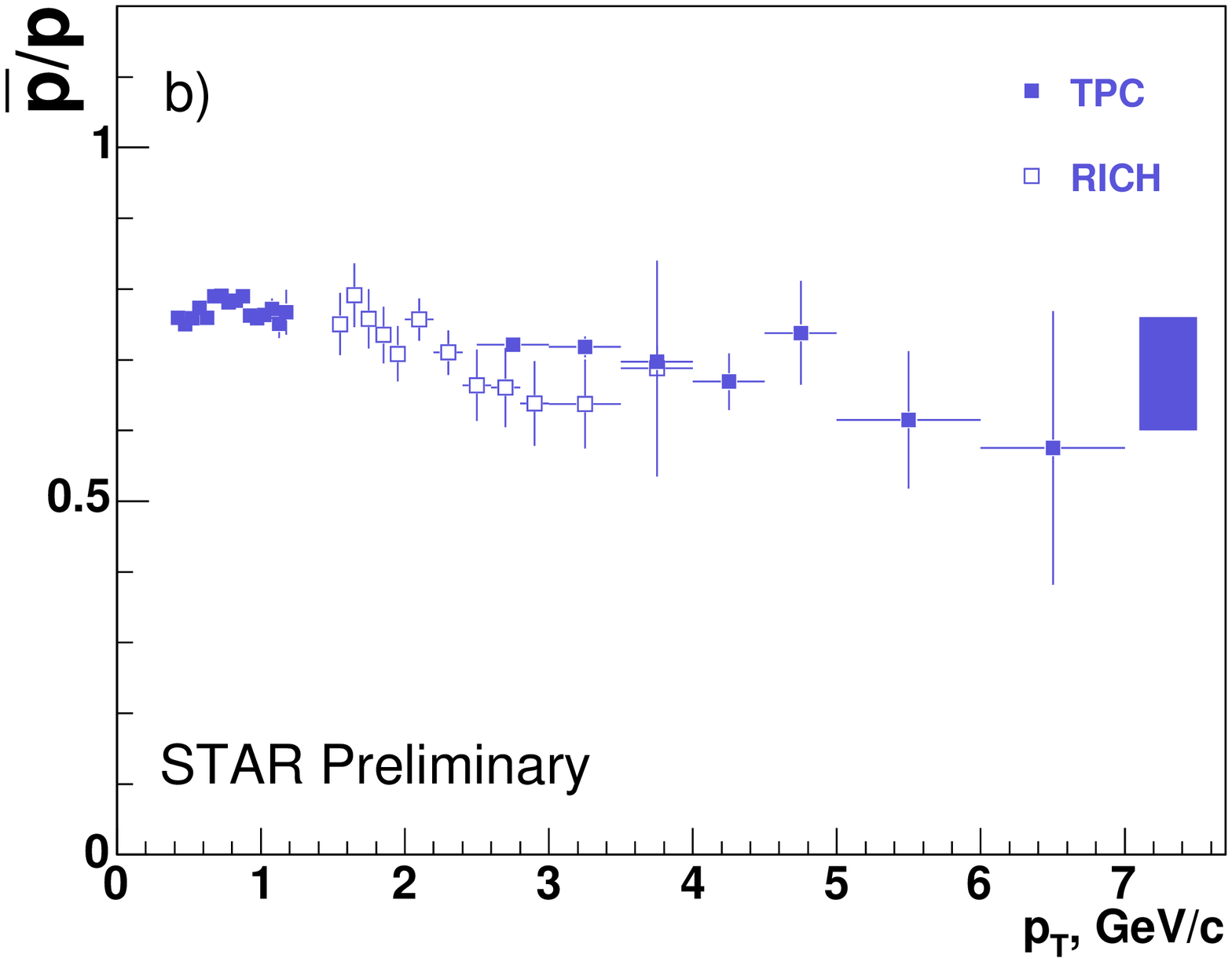}}
\caption[]{$\pi^+/\pi^-$ (a) and $\bar{p}/p$  (b) ratios as function of $p_T$. $dE/dx$ measurements are shown in filled symbols, RICH results are presented with open symbols. Colored bars represent estimates of systematic uncertainties  for the $rdE/dx$ measurement.}
\label{fig7}
\end{figure*}

Obtained $p_T$ dependence of $\bar{p}/h$ and $p/h$ ratios provides more insights onto what become known as ``baryon-meson puzzle'' (we use charged hadron measurements $h=\frac{h^++h^-}{2}$ from \cite{highpt}). The results presented in Figure 6 confirm reported anomalously high production of protons and antiprotons in the intermediate momentum range. After reaching maximum at about 2 GeV/$c$ both ratios  then fall off at higher transverse momentum.  
The  $p/h$ ratio indicates almost equal amounts of baryons and mesons are produced in central collisions in the range $2< p_T < 4$GeV/$c$. This is again in contradiction to the pQCD calculations, that give $p/\pi \approx 0.1-0.2$ \cite{b16}.
Different theoretical models predict variety of non-fragmentation contribution for  baryon production in the intermediate momentum range.
Various mechanisms for baryon production such as gluon junctions (Soft+Quench)\cite{matt2}, recombination \cite{matt1,hwa} and quark coalescence \cite{matt3} result in  baryon over meson enhancement in those models. All those models also predict rather quick fall off of those contributions at higher $p_T$. Given the currently large systematic uncertainty on our measurements (mostly driven by the statistical error on the $K_0^s$ measurement) we are unable at this point to favor particularly any of those models. We anticipated reduction of both statistical and systematic  errors on our measurements with the undergoing processing of high statistics Au+Au data sample obtained during Run-4 at RHIC.

\begin{figure*}[hbt]
\centerline{\epsfxsize=0.45\textwidth\epsfbox[0 0 567 449]{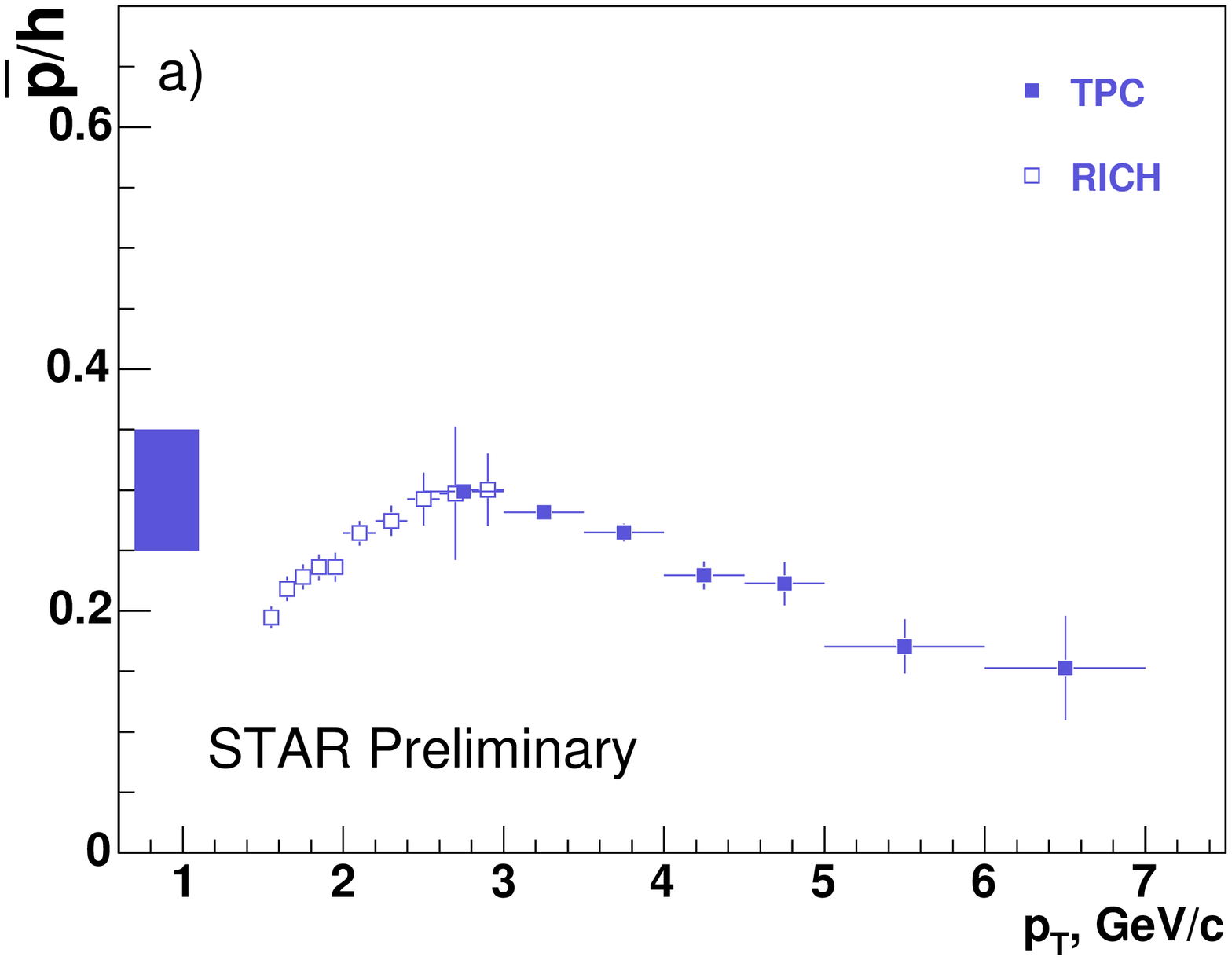}
\epsfxsize=0.45\textwidth\epsfbox[0 0 567 449]{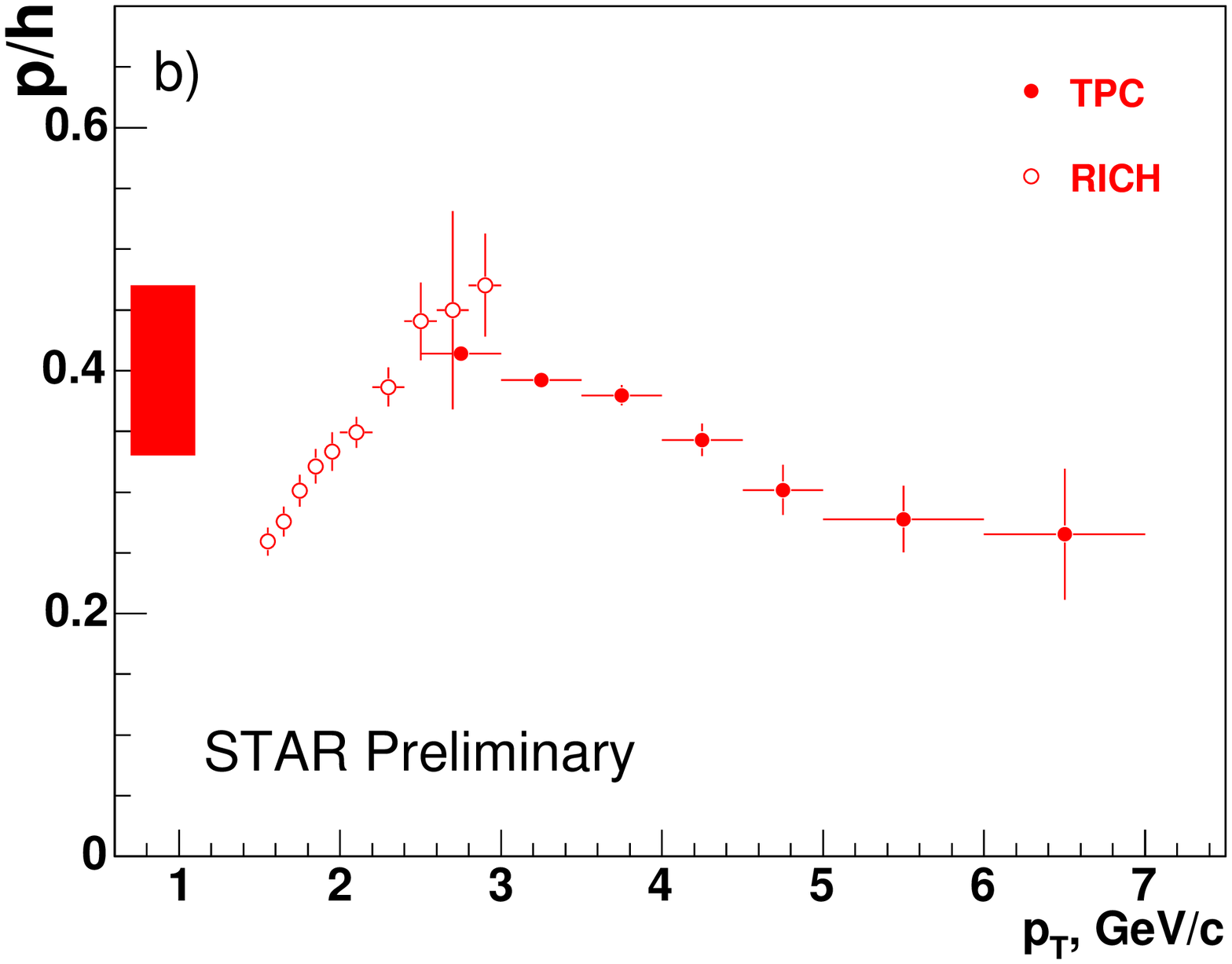}}
\caption[]{$\bar{p}/h$ (a) and $p/h$ (b) ratios as function of $p_T$. $dE/dx$ measurements are shown in filled symbols, RICH results are presented with open symbols. Colored bars represent estimates of systematic uncertainties  for $rdE/dx$ measurement.}
\end{figure*}

\section{Conclusions}\label{concl}

In summary, we have presented the first measurement of identified charged pions, protons and antiprotons at high transverse momentum from RHIC. Spectra and $p_T$ dependence of particle ratios have been presented for the momentum range of $2.5 <p_T< 7.0$~GeV/$c$. $\pi^-/\pi^+$ ratio shows no $p_T$-dependence in the momentum range covered and consistent with unity. The $\bar{p}/p$ measurements at intermediate and high transverse momentum found to be close within errors to the reported integrated yield ratios, dominated by low $p_T$ hadrons, and show little $p_T$ dependence.  We have  found that relative baryon to meson production has strong $p_T$ dependence: our measurement of both $\bar{p}/h$ and $p/h$ ratios show  smooth increase  as a function of transverse momentum at low $p_T$, reaching its maximum at about 2~GeV/$c$, followed by significant decrease in both ratios at higher transverse momentum.

\vfill\eject

\end{document}